# EFFICIENCY AND COMFORT OF KNEE BRACES: A PARAMETRIC STUDY BASED ON COMPUTATIONAL MODELLING

B.Pierrat[1], J.Molimard[2], P.Calmels[3], L.Navarro[4], S.Avril[5]

## 1. ABSTRACT

Knee orthotic devices are widely proposed by physicians and medical practitioners for preventive or therapeutic objectives in relation with their effects, usually known as to stabilize joint or restrict ranges of motion. This study focuses on the understanding of force transfer mechanisms from the brace to the joint thanks to a Finite Element Model. A Design Of Experiments approach was used to characterize the stiffness and comfort of various braces in order to identify their mechanically influent characteristics. Results show conflicting behavior: influent parameters such as the brace size or textile stiffness improve performance in detriment of comfort. Thanks to this computational tool, novel brace designs can be tested and evaluated for an optimal mechanical efficiency of the devices and a better compliance of the patient to the treatment.

## 2. INTRODUCTION

The knee is the largest joint in the body and is vulnerable to injury during sport activities and to degenerative conditions such as arthrosis. Knee injuries are common and account for 15-50% of all sports injuries [1]. Knee braces are prescribed for various knee syndromes such as ligament tears or disruptions, patellofemoral syndrome, iliotibial band syndrome, knee arthrosis and knee laxities [2]. These physio-pathological conditions involve pain and/or functional instability. These conditions are prevalent and are a huge burden on individuals and healthcare systems.

Numerous brace action mechanisms have been proposed and investigated such as proprioceptive improvements, strain decrease on ligaments, neuromuscular control enhancement, joint stiffness increase and corrective off-loading torque for unicompartimental knee osteoarthritis [3,4]. Studies aiming to justify the use of knee orthoses in medical practice were reviewed by [3–5]. The following conclusions have been reported:
  1. Mechanical/physiological effects have been highlighted, but their level and mechanisms remain poorly known.

---

[1] PhD Student, Ecole Nationale Supérieure des Mines, CIS-EMSE, CNRS:UMR5146, LCG, F-42023 Saint-Etienne, France

[2] Professor, Ecole Nationale Supérieure des Mines, CIS-EMSE, CNRS:UMR5146, LCG, F-42023 Saint-Etienne, France

[3] Doctor of Medicine, Laboratory of Exercise Physiology (LPE EA 4338), University Hospital of Saint-Etienne, 42055 Saint-Etienne CEDEX 2

[4] Research Associate, Ecole Nationale Supérieure des Mines, CIS-EMSE, CNRS:UMR5146, LCG, F-42023 Saint-Etienne, France

[5] Professor, Ecole Nationale Supérieure des Mines, CIS-EMSE, CNRS:UMR5146, LCG, F-42023 Saint-Etienne, France

2. Only a few high-level clinical studies exist, and the effectiveness of bracing versus no bracing on postoperative outcomes has not been conclusively demonstrated.

Possible explanations of 1 having no perceptible effect on 2 are that mechanical action levels are too low, or that patients do not comply to the orthopedic treatment and do not wear enough the device due to comfort issues. What is more, these studies are based on questionable methods and results lack authority. As a consequence of these uncertainties, medical practitioners and manufacturers still lack a simple evaluation tool for knee orthoses. A french committee of experts highlighted this problem [6] and stated that orthoses must be evaluated by taking both the mechanisms of action and the desired therapeutic effects into account.

In order to answer these issues, an original Finite Element Model approach has been developed. This model was built in agreement and cooperation with medical practitioners and orthotic industrials, in a tentative of linking design problems, brace ability to prevent a given pathology and patient comfort. As there are a huge variety of orthoses on the market, the focus was placed on mass-produced knee braces, in opposition to individualized orthotic devices. They are usually made of synthetic textiles and may incorporate bilateral hinges and bars, straps, silicone anti-sliding pads and patella hole. Different hinge systems exist in order to reproduce knee kinematics. A typical design of a usual brace is depicted in Fig. 1(a). They are prescribed either for prophylactic or functional purposes.

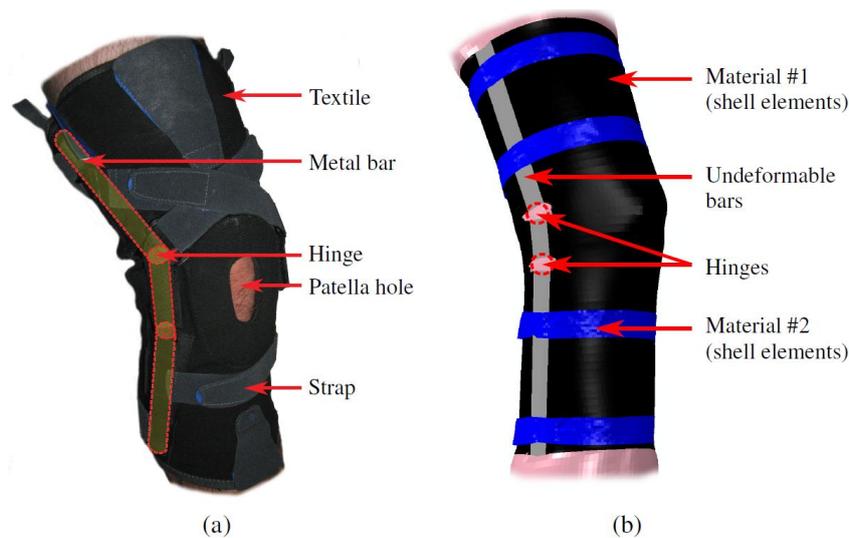

Fig. 1: Mass-produced knee orthosis: usual commercially available model (a) and FE model (b).

3. METHODS

3.1 Finite element model of the braced knee

The model was developed under Abaqus® v6.10-2. This generic model is not aimed to be patient specific, but to understand the force transfer mechanisms between the rigid parts of the knee brace and the joint through the brace fabric, the patient skin and soft tissues. 3D geometry of the human leg was obtained from a segmented PET-CT (Positron Emission Tomography - Computed Tomography) scan. The leg was scaled in order to reach the dimensions of a median French male leg (2006 French Measurement Campaign). It features undeformable bones, homogenized soft tissues (muscles, fat, tendons and fascias), skin and a fitted brace, as depicted in Fig. 1(b).

Soft tissues were meshed with 160 000 quadratic tetrahedral elements. The material was defined as homogeneous, isotropic, quasi-incompressible and hyper-elastic. A Neo-Hookean strain energy function was used as described in [7,8]. The model parameters for the leg were identified by [7] for a passive muscle ($G$ = 3-8 kPa) and by [8] for a contracted muscle ($G$ = 400 kPa). $K$ was set to 10×$G$ in order to enforce quasi-incompressibility.

The skin was meshed with 11 000 quadrilateral shell elements of thickness 1mm, as already modeled by [9]. The material was defined as homogeneous, isotropic, quasi-incompressible and hyper-elastic. An Ogden strain energy function was used as described in [9]. Values of $\alpha$ and $\mu$ have been identified by [9] on the forearm ($\alpha$ = 35 and $\mu$ = 15kPa). A pre-stress of 4 kPa was applied in circumferential and longitudinal directions of the skin at the start of the analysis.

Regarding the orthosis, the textile consisted of 30 000 quadrilateral shell elements and each strap of 1600 quadrilateral shell elements. The bars were modeled as rigid bodies. Mechanical behavior of fabrics has been already successfully modeled using shell elements [10]. The material was defined as homogeneous, orthotropic and linear elastic. The constitutive equations, written in vectorial form, relative to the warp and weft directions, are then:

$$\begin{pmatrix} N_{11} \\ N_{22} \\ N_{12} \end{pmatrix} = \begin{pmatrix} \frac{E_1}{1-\nu_{12}\nu_{21}} & \frac{\nu_{21}E_1}{1-\nu_{12}\nu_{21}} & 0 \\ \frac{\nu_{12}E_2}{1-\nu_{12}\nu_{21}} & \frac{E_2}{1-\nu_{12}\nu_{21}} & 0 \\ 0 & 0 & G_{12} \end{pmatrix} \begin{pmatrix} \epsilon_{11} \\ \epsilon_{22} \\ 2\epsilon_{12} \end{pmatrix} \quad (1)$$

and

$$\begin{pmatrix} M_{11} \\ M_{22} \\ M_{12} \end{pmatrix} = \begin{pmatrix} F_1 & \mu_2 F_1 & 0 \\ \mu_1 F_2 & F_2 & 0 \\ 0 & 0 & \tau_{12} \end{pmatrix} \begin{pmatrix} \kappa_{11} \\ \kappa_{22} \\ 2\kappa_{12} \end{pmatrix} \quad (2)$$

where $N_{ij}$ and $M_{ij}$ are the tensions and bending moments of the fabric, $\varepsilon_{ij}$ and $\kappa_{ij}$ the strains and bending strains, $E_i$ the tensile rigidities, $G_{12}$ the shear rigidity, $\nu_{ij}$ the Poisson's ratios, $F_i$ the bending rigidities, $\tau_{12}$ the torsional rigidity and $\mu_i$ parameters analogous to Poisson's ratios. Tensile rigidities and Poisson's ratios were obtained from unidirectional tensile tests on an Instron® machine whereas bending rigidities were measured using a KES-F device (Kawabata Evaluation System for Fabrics) [10,11]. Samples were taken from 4 commercially available orthoses and their straps.

Undeformable bars of the orthosis were connected using hinge connectors with a blocking feature, allowing them to pivot with the joint but not in the other way. A basic Coulomb friction model was used for the orthosis/skin and skin/soft tissues contacts in which contact pressure is linearly related to the equivalent shear stress with a constant friction coefficient $\mu$. Values of $\mu_{brace}$ for different fabric/skin systems are available in the literature, ranging from 0.3 to 0.7 [12,13]. Concerning the skin/soft tissues contact, no data was found in the literature for friction coefficient measurements. This parameter $\mu_{leg}$ was assumed to be 0.1. Skin was attached to soft tissues at the top and bottom of the leg.

A quasi-static analysis was performed using the Explicit solver. A joint kinematic was imposed, either a 20 mm front drawer, a 15° varus, a 20° pivot or a 45° flexion. A single analysis completed in about 4 hours on 8 CPUs at 2.4 Ghz.

3.2 Design Of Experiments

In order to characterize and grade the influence of brace design characteristics and patient-related specificities, 8 key parameters were identified. Their ranges or levels were chosen in agreement with existing brace designs and from data available in the literature. These parameters are detailed in Tab. 1.

| N° | Parameter | Associated manufacturer/ patient characteristic | Study range or levels |
|---|---|---|---|
| 1 | Tensile and bending stiffness of the fabric | Thread type and weaving technique of the fabric | $E_1 = E_2 = 200 \to 1000$ N/m |
| 2 | Initial radius of the brace | Brace size | $35 \to 70$ mm |
| 3 | Brace length | Brace length | 250 / 350 / 500 mm |
| 4 | Tensile and bending stiffness of the straps | Strap material | $E = 200 \to 25000$ N/m |
| 5 | Initial stress in the strap | Strap tightening | $200 \to 25000$ N/m |
| 6 | Strap shape | - | Parallel horizontal straps / Helical straps |
| 7 | Brace/skin friction coefficient $\mu_{brace}$ | Patient's skin humidity, anti-sliding interface material | $0.15 \to 0.5$ |
| 8 | Soft tissues stiffness | Muscle contraction | 5.5 / 400 kPa |

Tab. 1: Identified parameters and their area of study.

After normalizing the factors to a [-1;1] interval, 100 numerical simulations were chosen using an 8 dimension stratified latin hypercube sampling, which authorizes both continuous factors and given levels. 4 responses were output from the simulations: the slope of the reaction force/moment vs displacement/rotation curve for the drawer, varus and pivot (stiffness of the orthosis in a given direction) and the average contact pressure applied by the brace on the skin at the end of the flexion step. With the intention of comparing the parameters, the responses were normalized in such a way that their standard deviation was 0.5 and their mean 0. Finally, a linear regression was performed to find a first order, no-interaction polynomial response surface. The linearized effect of each parameter is the corresponding polynomial coefficient. A Fisher test with 91 degrees of freedom was used to determine how significant each factor is.

4. RESULTS

Before normalization, the results ranged as (mean ± standard deviation): drawer stiffness (1.64 ± 1.15 N/mm), varus stiffness (0.30 ± 0.28 N.m/°), pivot stiffness (0.053 ± 0.041 N.m/°) and average contact pressure (498 ± 357 Pa). The results of the parametric study are depicted in Fig. 2. The influence of each factor depends on the mechanical load, although the initial brace radius is a key parameter in each case. Other rather influent parameters were the fabric stiffness and the muscle contraction. Almost non-influent parameters were identified such as the strap shape and the friction coefficient. The influence of remaining parameters depends on the response. It is noteworthy that the contact pressure response is opposed to the stiffness responses, showing that a stiffer orthosis is also less comfortable. This trend is wrong for three parameters: the brace length, the strap stiffness and the strap tightening.

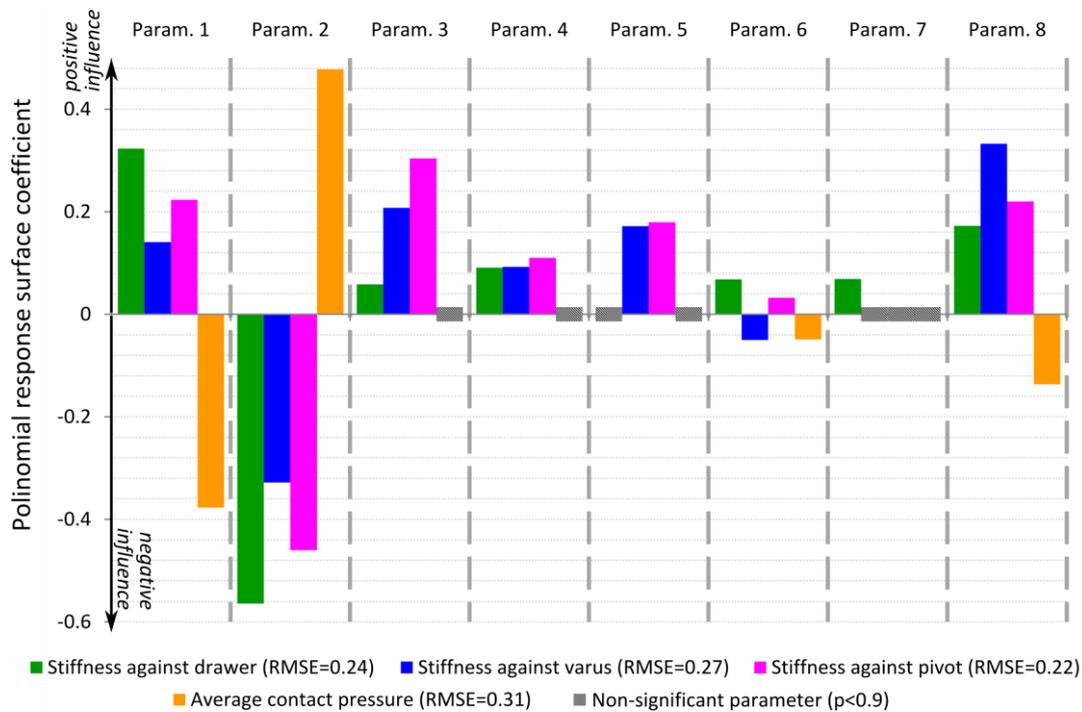

Fig. 2: Comparison of the effect of each parameter for different response surfaces.

5. DISCUSSION

The FE model is subject to limitations because it is not patient-specific. Even if it has not been demonstrated, it is highly probable that several patient-specific leg factors influence the mechanical response of the brace-leg system. These factors may include mechanical properties of the different leg constituents (skin, soft tissues) as well as the quantity of adipose tissue or the geometry of the leg itself. Nevertheless, the purpose of this work is not to compute the actual efficiency and comfort of a particular brace-leg system, but to understand the general mechanical mechanisms governing these phenomena. In that way, the developed generic model is perfectly suited, even if work remains to be done in validating and exploiting it. Besides, modeling choices may be subject to caution as most mechanical properties and friction models are derived from literature. Regarding the Design Of Experiments, the Root Mean Square Error (RMSE) of the linear response surface is quite high, indicating that the actual responses are probably not linear and that interactions between parameters exist. More FE simulations are required in order to compute such response surfaces with good reliability. Finally, the exploitation of the outcomes of this study indicates that manufacturers should focus on brace length and straps in order to increase joint stiffening without altering the brace comfort.

6. CONCLUSION

An adaptable FE model was successfully developed and used in a Design Of Experiments approach. Results showed that joint stiffening of knee braces may be increased by adjusting mechanically influent design parameters but caution must be exercised as brace stiffening results, in most cases, in an increase in discomfort. Only brace length and strap-related parameters efficiently stiffen the joint without altering comfort. Future work consists in validating the FE results using experimental means and developing an optimization method to contribute to the design of optimized orthoses.